\documentclass[aps,prl,twocolumn,superscriptaddress]{revtex4}
\usepackage{graphicx}
\usepackage{latexsym}
\usepackage{amssymb}
\usepackage{amsmath}
\usepackage{amsfonts}
\usepackage{bm}
\usepackage{multirow}
\usepackage{color}

\newcommand{\vect}[1]{{\bm{#1}}}
\newcommand{\eqnref}[1]{Eq.\,\eqref{#1}}
\newcommand{\figref}[1]{Fig.\,\ref{#1}}

\newcommand{\beq}{\begin{equation}}
\newcommand{\eeq}{\end{equation}}
\newcommand{\beqn}{\begin{eqnarray}}
\newcommand{\eeqn}{\end{eqnarray}}

\begin{document}

\title{Many-Body Localization of Symmetry Protected Topological States}



\author{Kevin Slagle}

\author{Zhen Bi}

\author{Yi-Zhuang You}

\author{Cenke Xu}

\affiliation{Department of physics, University of California,
Santa Barbara, CA 93106, USA}

\begin{abstract}

We address the following question: Which kinds of symmetry
protected topological (SPT) Hamiltonians can be many-body
localized? That is, which Hamiltonians with an SPT ground state
have finite energy density excited states which are all localized
by disorder? Based on the observation that a finite energy density
state, if localized, can be viewed as the ground state of a local
Hamiltonian, we propose a simple (though possibly incomplete) rule
for many-body localization of SPT Hamiltonians: If the ground
state and top state (highest energy state) belong to the same SPT
phase, then it is possible to localize all the finite energy
density states; If the ground and top state belong to different
SPT phases, then most likely there are some finite energy density
states which can not be fully localized. We will give concrete
examples of both scenarios. In some of these examples, we argue
that interaction can actually ``{\it assist}" localization of
finite energy density states, which is counter-intuitive to what
is usually expected.

\end{abstract}


\maketitle


\section{ 1. Introduction}

Symmetry protected topological (SPT) states and many-body
localization (MBL) are two striking phenomena of quantum many-body
physics. A $d-$dimensional SPT state is the ground state of a
local Hamiltonian whose $d-$dim bulk is fully gapped and
nondegenerate, while its $(d-1)-$dim boundary is gapless or
degenerate when and only when the system preserves a certain
symmetry $G$~\cite{wenspt,wenspt2}. An SPT state must have ``short
range entanglement"; meaning that the entanglement entropy of its
subsystems scales strictly with the area of the boundary of the
subsystem: $\mathcal{S}_A \sim L^{d-1}$~\cite{bauer}, where $L$ is
the linear size of the subsystem $A$. MBL refers to a phenomenon
of the entire spectrum of a local Hamiltonian with disorder,
including all of the highly excited states with finite energy
density. Localization of single particle states under quenched
disorder is well-understood~\cite{anderson}, and recent studies
suggest that localization can survive under
interaction~\cite{atshuler,polyakov}.
In our current work the phrase MBL refers to systems whose all
many-body eigenstates are localized, namely the entanglement
entropy of all finite energy density states obey the same area law
as SPT states instead of the usual volume law typically obeyed by
finite energy density states.

These observations imply that in a many-body localized system, any
finite energy density state actually behaves like the ground state
of a local parent Hamiltonian. Indeed, it was proposed that
phenomena such as stable edge states and spontaneous symmetry
breaking~\cite{huse,sondhi,ashvinaltman,bauer}, which usually
occur at the ground state of a system, can actually occur in
finite energy density states of MBL systems. In fact, we can {\bf
define a MBL system as a system for which any finite energy
density eigenstate is a short range entangled ground state of a
local parent Hamiltonian}. And if the system preserves a certain
symmetry, {\bf then any finite energy density state of the MBL
system should also obey the classification of SPT states}. Then we
can view energy density $\varepsilon$ as a tuning parameter
between SPT states. Of course, in the thermodynamic limit, because
there are infinite states in an infinitesimal energy density
interval $(\varepsilon, \varepsilon + d\varepsilon)$, we expect
there exists many $1d$ curves in the spectrum parameterized by
$\varepsilon$ with one state $|\psi\rangle_\varepsilon$ at each
$\varepsilon$, which is the ground state of an effective SPT
Hamiltonian $H_{\varepsilon}$. And on each such curve
$|\psi\rangle_\varepsilon$ is (roughly speaking) continuous in the
sense that $|\psi\rangle_\varepsilon$ and
$|\psi\rangle_{\varepsilon + d \varepsilon}$ are similar (despite
being orthogonal), namely physical quantities averaged over the
entire system change continuously with $\varepsilon$ on this
curve. In an ergodic system, the eigenstate thermalization
hypothesis~\cite{eth} implies that most states with similar energy
density $\varepsilon$ are similar (their reduced density matrices
all behave like a thermal density matrix); in a MBL state,
although states with the same energy density can in principle be
very different, we still expect (assume) that the continuous
curves mentioned above exist, although states in different curves
can be very different.

Within one of these curves mentioned above, tuning $\varepsilon$
is just like tuning between the ground states of local
Hamiltonians. Furthermore, by tuning $\varepsilon$ there may or
may not be a phase transition. In particular, if all excited
states belong to the same SPT phase for arbitrary energy density
$\varepsilon$, then there does not have to be any quantum phase
transition when tuning $\varepsilon$, which implies that all of
the excited states have short range correlations and area-law
entanglement entropy, $i.e.$ all the finite energy density states
are localized; on the other hand, if states with different energy
density $\varepsilon$ on the same curve belong to different SPT
phases, then there must be at least one phase transition at
certain critical energy on this curve when tuning $\varepsilon$.
This phase transition behaves just like an ordinary zero
temperature quantum phase transition between different quantum
ground states under disorder. For $1d$ systems this ``critical"
energy density state could be in the ``infinite-randomness"
phase~\cite{ma1,ma2,danfisher1,danfisher2}, whose entanglement
entropy scales logarithmically with the subsystem
size~\cite{moorerefael}, hence it is not fully localized. The
existence of the ``infinite-randomness" states at finite energy
density have already been observed in Ref.~\onlinecite{sid}.

Due to the fact that in a generic nonintegrable Hamiltonian $H$,
the ground state $| G \rangle$ and top state $| T \rangle$
(highest energy state of $H$ and also ground state of $-H$) are
usually the easiest states to analyze, the most convenient way to
determine the existence of ``critical" states in the spectrum is
to check whether the ground and top states belong to the same SPT
phase or not. In summary, if $|G\rangle$ and $|T\rangle$ belong to
different SPT phases, and if we understand that these two SPT
states are separated by one or multiple continuous phase
transitions (this will depend on the type of SPT phases
$|G\rangle$ and $|T\rangle$ belong to), then there must be some
``critical" excited states in the spectrum which cannot be fully
localized~\footnote{In this work, the phrase ``SPT states" also
include direct product states, we view direct product states as
``trivial" SPT states. So far not all quantum phase transitions
between SPT states have been completely studied, and it is
possible that some SPT states are separated by a first order
transition. Our statement only applies to the cases that
$|G\rangle$ and $|T\rangle$ belong to two different SPT phases
that we know are separate by a continuous phase transition, for
example the transition between the topological superconductor and
the trivial state of the Kitaev's chain (section 2B).}. We will
apply this rule to various examples in the next section.




\section{2. Examples}

\subsection{ 2A. Kitaev's chain: localization}

We first apply our argument to the Kitaev's chain:
\begin{align}
H = \sum_j - \left( t + (-)^j \delta t + \Delta t_j \right) i
\gamma_j \gamma_{j+1}, \label{kitaev1}
\end{align}
where $\gamma_j$ are Majorana fermions and $\Delta t_j$ is a
random hopping parameter with zero mean and standard deviation
$\sigma_{\Delta t}$. The topological superconductor phase ($\delta
t > 0$) and the trivial phase ($\delta t < 0$) can both be fully
localized by disorder, because for either sign of $\delta t$, the
ground state $|G\rangle$ and top state $|T\rangle$ both belong to
the same phase (we choose the convention that $(2j-1, 2j)$ is a
unit cell). This can be seen in the clean limit with $\Delta t =
0$. In momentum space $H = \sum_k d^x(k) \tau^x + d^y(k)\tau^z$,
and $\vec{d}$ is a nonzero O(2) vector in the entire $1d$
Brillouin zone with $\delta t \neq 0$. For either sign of $\delta
t$, $H$ and $-H$ have the same topological winding number $n_1 =
\frac{1}{2\pi} \int dk \ \hat{d}^a \partial_k \hat{d}^b
\epsilon_{ab}$; thus $|G\rangle$ and $|T\rangle$ belong to the
same phase. Based on our argument, all the finite energy states
with either sign of $\delta t$ can be fully localized by random
hopping $\Delta t$. The only states not fully localized in the two
dimensional phase diagram tuned by $\varepsilon$ and $\delta t$
are located at the critical line $\delta t = 0$. The critical line
$\delta t = 0$ is in a ``infinite-randomness" fixed point, and it
can be understood through the strong disordered real space
renormalization group~\cite{danfisher1,danfisher2,moorerefael,RSRGX1,RSRGX2,sid}.

\begin{figure}[htbp]
\begin{center}
\includegraphics[width=\columnwidth]{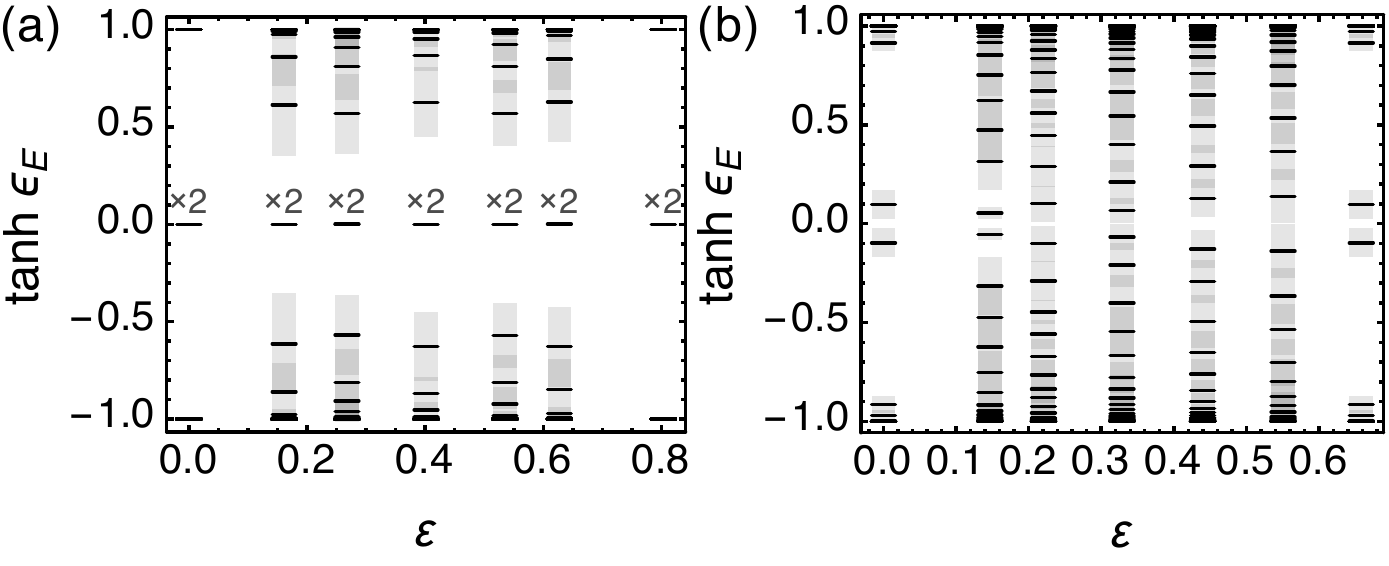}
\caption{Single-particle entanglement spectrum for many-body
eigenstates of the random Kitaev's chain, at (a) $\delta t = 0.5
t$ and (b) $\delta t = 0$. In both cases $\sigma_{\Delta t}=0.3
t$. We take a 128-site system with periodic boundary condition,
which is partitioned into two 64-site subsystems for the
entanglement calculation. $\epsilon_E$ is the single-particle
entanglement energy (s.t. the reduced density matrix
$\rho_A=\exp(-c^\dagger \epsilon_{E} c)$, as shown in
Ref.~\onlinecite{frank}). The spectrum of $\epsilon_E$ is shown as
$\tanh\epsilon_E$, and is calculated for several many-body
eigenstates: including the ground and the top states and other 5
randomly picked finite energy density states, which are arranged
in order of their energy density $\varepsilon$. The shading
denotes the standard deviation of the entanglement energy levels
under a disorder average over the system. Of note are the
topologically non-trivial, two-fold degenerate, zero energy modes
throughout the entire spectrum $\varepsilon$ in the topological
phase (a).} \label{fig: Kitaev entanglement}
\end{center}
\end{figure}

Here we confirm the conclusions in
Ref.~\onlinecite{huse,ashvinaltman,bauer} that the finite energy
density excited states of the Kitaev's chain with $\delta t
> 0$ are still ``topological". Since the energy level spacing
between two eigenstates vanishes in the thermodynamic limit, the
best way to determine if an excited state is topological or not is
to compute its entanglement spectrum (the system is defined on a
periodic 1d lattice). And because the system is noninteracting, we
will compute the single-particle entanglement spectrum introduced
in Ref.~\onlinecite{frank} for each excited state. The
single-particle entanglement spectrum for the topological phase
($\delta t > 0$) is shown in \figref{fig: Kitaev entanglement}(a),
where two zero energy modes can be observed in the spectrum
(corresponding to the Majorana zero modes at both entanglement
cuts respectively). This topologically non-trivial feature
persists for all energy eigenstates in the many-body spectrum,
including the ground/top states and the finite energy density
states in between. However at the critical line $\delta t =0$, as
shown in \figref{fig: Kitaev entanglement}(b), the zero energy
modes are lifted by the long-range entanglement, and the
single-particle entanglement levels become gapless around
$\epsilon_E=0$ 
which leads to the logarithmic scaling of the entanglement
entropy.

The Kitaev's chain itself is just a free fermion model. But our
argument indicates that under interaction, as long as $|G\rangle$
and $|T\rangle$ are still both in the topological superconductor
phase, all of the excited states can still be localized. Such a
generalization is justified given that the non-interacting
Anderson localized states can be adiabatically connected to the
many-body localized states under interaction, as proven in
Ref.\,\cite{bauer}.

\subsection{2B. Modified Kitaev's chain: critical states and
interaction assisted localization}

\begin{figure}
\includegraphics[width=.45\textwidth]{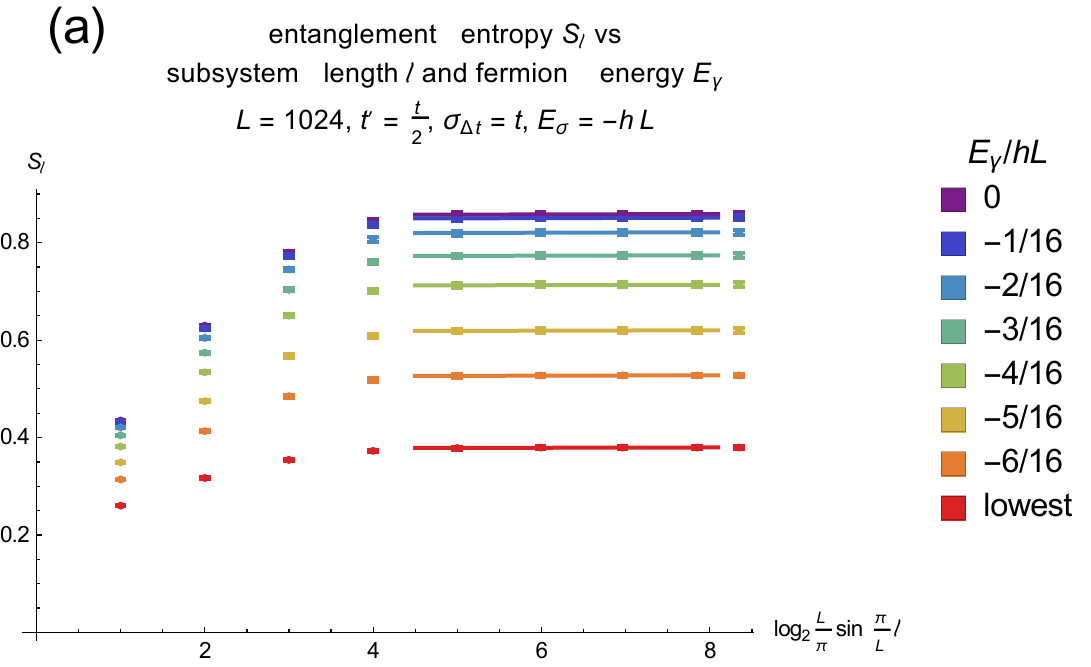}
\includegraphics[width=.45\textwidth]{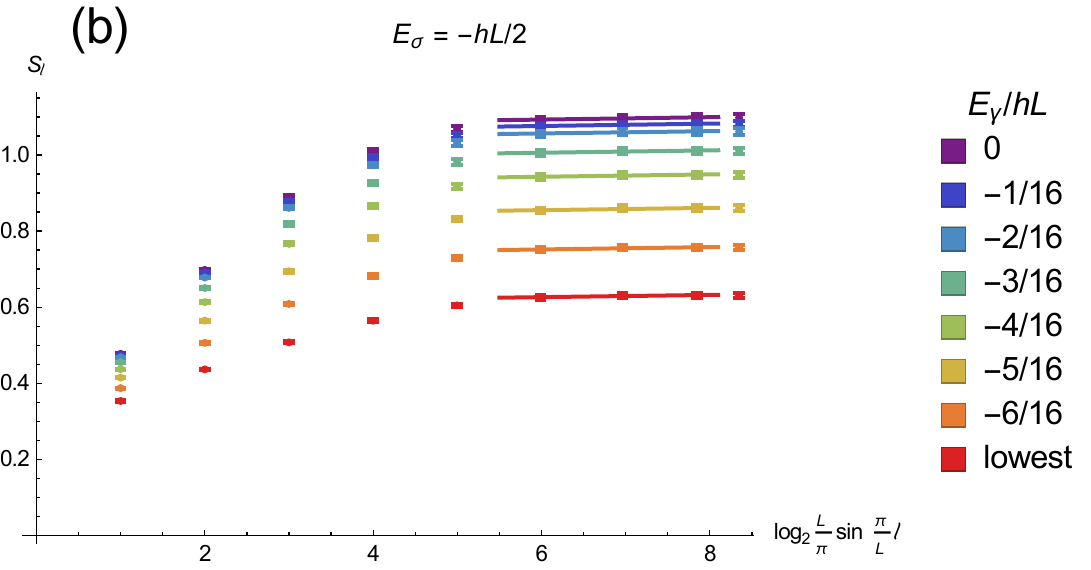}
\includegraphics[width=.45\textwidth]{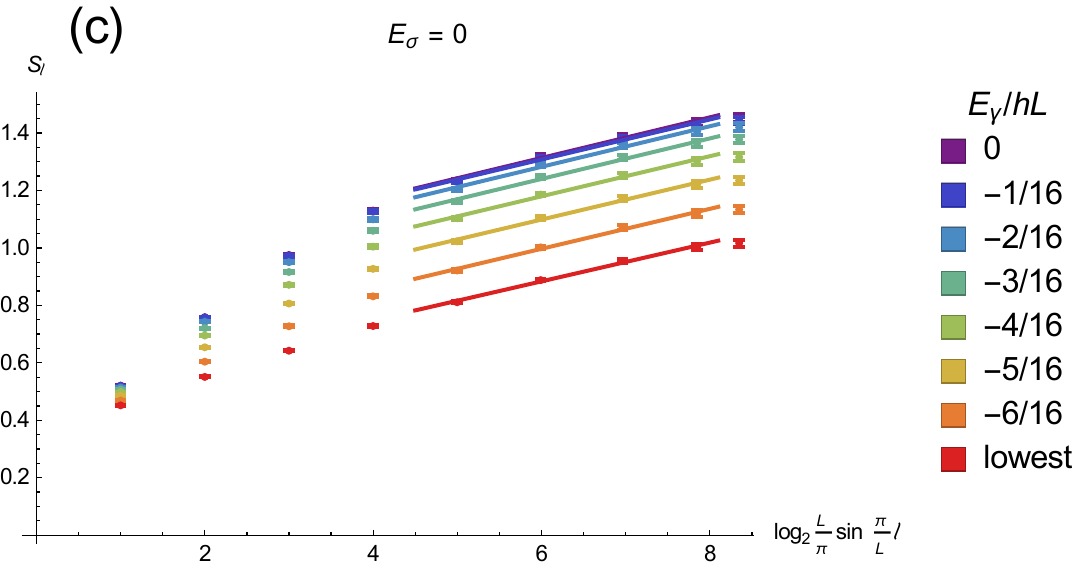}
\caption{Entanglement entropy $S_\ell$ vs log subsystem length
$\log_2\ell$ vs fermion energy $E_\gamma$ (energy of the first
term in \eqnref{kitaev2}) for various boson energies $E_\sigma
\equiv -h\sum_j\sigma_j^z= -h L, -h L/2, 0$ (a,b,c) (second term
in Eq.~\ref{kitaev2}). Calculations are done on a random Majorana
chain with $L=1024$ sites, and the standard deviation of $\Delta
t_j$ is $\sigma_{\Delta t} = t$. States with $E_\sigma = 0$ are
the critical excited states which are delocalized. All states with
different $E_\gamma$ at $E_\sigma = 0$ have logarithmic
entanglement entropy, and hence are delocalized.}\label{EEvsL}
\end{figure}

In this subsection we consider a modified Kitaev's chain:
\begin{align}
H = \sum_j - \left( t - (-1)^j t^\prime  \sigma^z_j + \Delta t_j
\right) i \gamma_j \gamma_{j+1} - h \sigma^z_j, \label{kitaev2}
\end{align}
where again $\Delta t_j$ is random and $t, t^\prime, h > 0$. In
this model $\sigma^z_j$ commutes with the Hamiltonian, which
implies that any energy eigenstate will also be an eigenstate of
$\sigma^z_j$. In the clean limit, the ground state $|G\rangle$ of
the system has $\sigma^z_j = 1$ everywhere, and the fermions are
in the trivial phase; in contrast, $|T\rangle$ must have
$\sigma^z_j = -1$ everywhere, and hence $|T\rangle$ is in the
topological superconductor phase. With disorder, both states can
be localized, and their entanglement entropy shows the area-law
scaling (\emph{i.e.} $\mathcal{S}\sim \text{const.}$ for $1d$) as
in \figref{EEvsL}(a). But since the ground state and the top state
belong to different SPT phases, based on our argument, there must
be some finite energy density states which cannot be fully
localized. In this model it is easy to visualize these delocalized
excited states. An excited state of the system has a static
background configuration of $\sigma^z_j$ which does not satisfy
$\sigma^z_j = 1$. If we consider a random configuration of
$\sigma^z_j$ that has the average $\overline{\sigma^z_j} = 0$,
then one can simply absorb $\sigma^z_j$ into the random numbers
$\Delta t_j$, and the effective Hamiltonian for Majorana fermions
$\gamma_{j}$ reads $H_\mathrm{eff} = \sum_j - \left( t + \Delta
t^\prime_j \right) i \gamma_j \gamma_{j+1}$, which is precisely
the random hopping Majorana fermion model \eqnref{kitaev1} tuned
to the critical point $\delta t =  0$. And according to
Ref.~\onlinecite{danfisher2,moorerefael}, the ground state of
$H_{\mathrm{eff}}$ (which is a highly excited state of the
original Hamiltonian \eqnref{kitaev2} due to the $h$ term) has a
power-law correlation after disorder average, and its entanglement
entropy scales logarithmically with the subsystem size:
$\mathcal{S} \sim \log \ell$~\cite{moorerefael}. So the
delocalization happens right at the energy scale $E_\sigma\equiv
-h\sum_j\sigma_j^z =0$. In deed our numerical calculation shows
that as long as $E_\sigma\neq0$, the eigen states are all
localized with area-law entanglement entropy as in
\figref{EEvsL}(a,b); but for $E_\sigma=0$, the eigen states are
delocalized with logarithmically-scaled entanglement entropy as in
\figref{EEvsL}(c). Thus the model \eqnref{kitaev2} cannot be fully
many-body localized, which is consistent with our statement made
in the introduction.

The model Eq.~\ref{kitaev2} has a time-reversal symmetry $T:
\gamma_j \rightarrow (-)^j \gamma_j$ and $\sigma^z_j \rightarrow
\sigma^z_j$. It is known that with this time-reversal symmetry and
without interactions, the Kitaev's chain has $\mathbb{Z}$
classification~\cite{ludwigclass1,ludwigclass2,kitaevclass}; that
is with an arbitrary number of flavors of Eq.~\ref{kitaev2},
$|T\rangle$ is always a nontrivial topological superconductor,
while $|G\rangle$ is always a trivial phase. However under certain
flavor mixing four-fermion
interaction~\cite{fidkowski1,fidkowski2}, the classification of
Kitaev's chain with time-reversal symmetry reduces to
$\mathbb{Z}_8$. Namely under this four-fermion interaction, for
eight copies of Eq.~\ref{kitaev2}, $|G\rangle$ and $|T\rangle$
become the same trivial phase, which implies that there does not
have to be any phase transition when increasing $\varepsilon$, and
all of the finite energy density excited states can be fully
localized under the interplay between disorder and interaction.

In model Eq.~\ref{kitaev2}, the logarithmic entanglement entropy
at the critical excited state comes from the long range effective
hopping under renormalization
group~\cite{danfisher1,danfisher2,moorerefael}. We can assume that
the four-fermion interaction on each site is random,
then when and only when there are $8k$ copies of
Eq.~\ref{kitaev2}, under interaction each site independently
possesses a random set of many-body spectrum {\it without
degeneracy}. Let $\delta V$ be the typical energy level spacing of
the interaction Hamiltonian on each site. To create entangled
pairs between distant sites, the effective long-range coupling
$t_\text{eff}$ generated under RG must overcome the energy scale
of $\delta V$ to hybridize the many-body states. However the
effective coupling strength actually falls rapidly with the
distance\cite{danfisher1,danfisher2, moorerefael} as
$t_\text{eff}\sim t e^{-\sqrt{r}}$, so the long-range coupling can
only lead to exponentially small entanglement
$\Delta\mathcal{S}\sim(t_\text{eff}/\delta V)^2\sim(t/\delta
V)^2e^{-2\sqrt{r}}$. Therefore even with {\it weak} interaction,
all of the eigenstates are short-range entangled area-law states,
and can be fully localized. In contrast, without interaction, no
matter what kind of fermion-bilinear perturbations we turn on in
Eq.~\ref{kitaev2}, as long as these terms preserve the
time-reversal symmetry defined above and the topological nature of
$|G\rangle$ and $|T\rangle$, there must {\it necessarily} be some
finite energy density states which cannot be fully localized. Thus
in this case {\it interaction actually ``assists" many-body
localization}, which is opposite from what is usually expected for
weak interaction, in for example Ref.~\onlinecite{potter}, and is
also different from the strong interaction reinforced localization
studied in
Ref.~\onlinecite{interaction1,interaction2,interaction3}.

Notice that this ``interaction assisted localization" is only
possible with $8k$ copies of the Kitaev's chain with time-reversal
symmetry. With 4 copies of the Kitaev's chain, the spectrum on
each site contains two sets of two-fold degenerate states even
under interaction that preserves time-reversal, then the effective
long-range coupling $t_\text{eff}$ generated under RG will still
lead to maximal entanglement between distant sites. A detailed RG
analysis about this will be given in another paper~\cite{youxuRG}.

\subsection{2C. Bosonic SPT states, Haldane phase}

Many bosonic SPT parent Hamiltonians can be written as a sum of
mutually commuting local terms. For example, the ``cluster model"
for the $1d$ SPT with $Z_2 \times Z_2$
symmetry~\cite{ashvinaltman}, the Levin-Gu model~\cite{levingu}
and the CZX model~\cite{czx} for the $2d$ SPT states with $Z_2$
symmetry, and the $3d$ bosonic SPT state with time-reversal
symmetry~\cite{fiona3d} are all a sum of commuting local
operators; thus their ground states are a product of eigenstates
of local operators~\footnote{Ref.~\onlinecite{wenspt} actually
proposed a general way of constructing parent Hamiltonians for all
bosonic SPT states within the group cohomology classification.
However, in Ref.~\onlinecite{wenspt} the local Hilbert space is
labeled by group elements, which implies that for a system with
continuous symmetry the local Hilbert space in
Ref.~\onlinecite{wenspt}'s construction already has infinite
dimension, and hence its excited states can also have infinite
local energy density. In this work we only discuss systems with a
finite dimensional Hilbert space and finite energy density.}. SPT
Hamiltonians written in this form are very similar to the
``universal" Hamiltonian of MBL state proposed in
Ref.~\onlinecite{dimaconserve}, which is also a sum of mutually
commuting local terms, because a MBL system has an infinite number
of local conserved quantities.

All of the idealized SPT models mentioned above have a $Z_2$
classification, and their ground and top states belong to the same
SPT phase. Obviously there should be no phase transition while
increasing energy density $\varepsilon$. This statement is still
valid with small perturbations which make these models
nonintegrable as long as the nature of $|G\rangle$ and $|T\rangle$
are not affected by the perturbations. Thus these models (and
their nonintegrable versions) can all be fully localized by
disorder.

However, some other bosonic SPT models can not be fully localized.
In the following we will give one such example for the Haldane
phase~\cite{haldane1,haldane2}: \begin{align} H = \sum_j (-1)^j (J
+ \Delta J_j) \vect{S}_{j} \cdot \vect{S}_{j+1} + \cdots
\end{align} $\vect{S}_{j}$ are spin-1/2 operators. The ellipsis
includes perturbations that break the system's symmetry down to a
smaller symmetry (such as time-reversal or $Z_2 \times Z_2$) that
is sufficient to protect the Haldane phase, but do not lead to
degeneracy in the bulk spectrum, namely only the boundary
transforms nontrivially under symmetry.
If the random coupling $\Delta J_j$ is not strong enough to change
the sign of $J$, then the ground state and top state of this model
correspond to two opposite dimerization patterns of the spin-1/2s.
Thus one of them is equivalent to the Haldane's phase while the
other is a trivial phase as long as we pick a convention of
boundary. If we assume the random Heisenberg coupling $\Delta J$
is sufficient to localize most of the excited states, then there
must be an unavoidable phase transition while increasing energy
density $\varepsilon$. According to our argument in the
introduction, this phase transition should behave just like an
ordinary quantum phase transition at zero temperature. It is known
that the quantum phase transition between a Haldane phase and a
trivial phase is a conformal field theory, and it is equivalent to
a spin-1/2 chain without dimerization. With strong disorder, this
quantum critical point will be driven into the infinite-randomness
spin singlet phase~\cite{ma1,ma2,danfisher1,danfisher2} with a
power-law decaying disorder averaged spin-spin correlation
function and a logarithmic entanglement
entropy~\cite{moorerefael}.

\subsection{2D. $2d$ interacting topological superconductor:
critical states and interaction assisted localization}

In this subsection we will discuss the nonchiral $2d$ $p\pm ip$
topological superconductor, $i.e.$ $p+ip$ pairing for spin-up
fermions, and $p-ip$ pairing for spin-down fermions. On a square
lattice this TSC can be written in the Majorana fermion basis:
\beqn H &=& \sum_k \chi^t_{-k} ( \tau^x \sin k_x + \tau^z \sigma^z
\sin k_y ) \chi_k \cr\cr &+& \chi_{-k}^t \tau^y (e - \cos k_x -
\cos k_y) \chi_k, \label{pip}\eeqn where $\sigma^z = \pm 1$
represents spin-up and down, while $\tau^z = \pm 1$ represents the
real and imaginary parts of the electron operator. Without any
symmetry, this system is equivalent to the trivial state, $i.e.$
its boundary can be gapped out without degeneracy. However, when $
0<e<2$, with a $Z_2$ symmetry which acts as $Z_2 : \chi
\rightarrow \sigma^z \chi$, the system is a nontrivial TSC. This
system can also have another time-reversal symmetry, which is
unimportant to our analysis. The boundary of this system reads: $
H = \int dx \ \chi^t ( - i
\partial_x \sigma^z) \chi$, $Z_2: \chi \rightarrow
\sigma^z \chi$. The $Z_2$ symmetry forbids any single particle
backscattering at the boundary for arbitrary copies of the system,
thus the $p \pm ip$ TSC with the $Z_2$ symmetry has a $\mathbb{Z}$
classification without interaction.

Without any interaction, for $n-$copies of the $p\pm ip$ TSC,
$|G\rangle$ and $|T\rangle$ belong to different SPT phases. This
is because for either spin-up or down fermions, the Chern number
of $|G\rangle$ and $|T\rangle$ are opposite. And because the
system has a $\mathbb{Z}$ classification, $|G\rangle$ and
$|T\rangle$ must belong to different SPT states. Using our
argument in the introduction, this implies that under disorder
that preserves the $Z_2$ symmetry, there must be some finite
energy density states which cannot be fully localized. This is not
surprising, considering that even at the single particle level
there are likely extended single particle states under disorder.
The existence of extended single particle states is well-known in
integer quantum Hall state~\cite{halperiniqh}, and recently
generalized to quantum spin Hall insulator with a $\mathbb{Z}_2$
index~\cite{localqsh1,localqsh2}.

The situation will be very different with interactions. Once again
a well-designed interaction will reduce the classification of this
$p \pm ip$ TSC from $\mathbb{Z}$ to
$\mathbb{Z}_8$~\cite{qiz8,yaoz8,zhangz8,levinguz8}. Namely
$n-$copies of Eq.~\ref{pip} is topologically equivalent to
$(n+8k)-$copies. This implies that under interaction $|G\rangle$
and $|T\rangle$ actually belong to the same phase when $n = 4k$.
Thus when $n = 4k$, the phase transition in the noninteracting
limit will be circumvented by interaction above a certain critical
value. Thus once again interaction {\it assists} MBL in this case.
When $n = 8$, $|G\rangle$ and $|T\rangle$ are both trivialized by
interaction, namely interaction can adiabatically connect both
states to a direct product of local states. When $n = 4$,
Ref.~\onlinecite{fb} showed that interaction can confine the
fermionic degrees of freedom, and drive four copies of the $p\pm
ip$ TSC into a $2d$ bosonic SPT state with $Z_2$ symmetry, which
as we discussed in the previous section, can also be fully
many-body localized.

Please note that in the noninteracting limit the quantum phase
transition between $2d$ TSC and trivial state is described by
gapless $(2+1)d$ Majorana fermions, and since a weak short range
four-fermion interaction is irrelevant for gapless $(2+1)d$
Dirac/Majorana fermions, only strong enough interaction can gap
out the quantum phase transition. Thus unlike the $1d$ analogue
discussed in section{\bf 2B}, we expect that in this $2d$ system
only strong enough interaction can ``assist" disorder and localize
all the excited states even for $n = 4k$.

\section{3. Summary}

In this work we propose a simple rule to determine whether a local
Hamiltonian with symmetry can be many-body localized. Since MBL is
a phenomenon for the entire spectrum, we need to start with a
lattice Hamiltonian for our analysis. Therefore the low energy
field theory descriptions and classification of SPT states such as
the Chern-Simons field theory~\cite{luashvin} and the nonlinear
sigma model field theory~\cite{xuclass} will not be able to
address this question. Instead, our argument is based on the
nature of the ground and top states of the same lattice
Hamiltonian. Our argument is general enough, that it can be
applied to both free and interacting systems, bosonic and
fermionic SPT systems. And counterintuitively, we found that
because interactions change the classification of fermionic
topological insulators and topological superconductors, in some
cases interactions actually assists localization, rather than
delocalization.

The authors are supported by the the David and Lucile Packard
Foundation and NSF Grant No. DMR-1151208. The authors are grateful
to Chetan Nayak for very helpful discussions.

\bibliography{SPTMBL}

\end{document}